\documentclass
{revtex4}%
\usepackage{amsmath}
\usepackage{color}
\usepackage{graphicx}
\begin{document}

\title[Nonlinear emission dynamics of a GaAs microcavity]{Nonlinear emission
dynamics of a GaAs microcavity with embedded quantum wells}

\author{V V Belykh}
\email{belykh@lebedev.ru}
\affiliation{P.N. Lebedev Physical
Institute, Russian Academy of Sciences, Moscow, 119991 Russia}
\author{V A Tsvetkov}
\affiliation{P.N. Lebedev Physical Institute, Russian Academy of
Sciences, Moscow, 119991 Russia}
\author{M L Skorikov}
\affiliation{P.N. Lebedev Physical Institute, Russian Academy of
Sciences, Moscow, 119991 Russia}

\author{N N Sibeldin}
\affiliation{P.N. Lebedev Physical Institute, Russian Academy of
Sciences, Moscow, 119991 Russia}

\begin{abstract}
The emission dynamics of a GaAs microcavity at different angles of
observation with respect to the sample normal under conditions of
nonresonant picosecond-pulse excitation is measured. At sufficiently
high excitation densities, the decay time of the lower-polariton
emission increases with the polariton wavevector; at low excitation
densities the decay time is independent of the wavevector. The
effect of additional nonresonant continuous illumination on the
emission originating from the bottom of the lower polariton branch
is investigated. The additional illumination leads to a substantial
increase in the emission intensity (considerably larger than the
intensity of the photoluminescence excited by this illumination
alone). This fact is explained in terms of acceleration of the
polariton relaxation to the radiative states due to scattering by
charge carriers created by the additional illumination. The results
obtained show, that at large negative detunings between the photon
and exciton modes, polariton--polariton and polariton--free carrier
scattering are the main processes responsible for the filling of
states near the bottom of the lower polariton branch.
\end{abstract}

\pacs{78.67.Pt, 78.67.De, 78.47.jd, 78.55.Cr}
\maketitle

\section{Introduction}
For almost two decades, studies of exciton polaritons in
semiconductor microcavities (MC) \cite{Weisbuch} attract
considerable attention inspired by a diversity of interesting
features in this system: polariton Bose--Einstein condensation (BEC)
\cite{Kasprzak,Balili,Christopoulos,Wertz,Roumpos}, superfluidity
\cite{Amo,Amo1}, stimulated polariton--polariton scattering
\cite{Savvidis,Stevenson,Tartakovskii,Krizhanovskii,Kulakovskii}
etc. (see book \cite{Kavokin} and references therein).

MC polariton energy relaxation is of particular interest, because it
has a profound effect on many properties of the polariton system. It
is well known that relaxation of MC polaritons to the bottom of the
lower polariton branch (LPB) is hampered by the bottleneck effect
\cite{Tassone}. It is the bottleneck effect together with the small
polariton lifetime near the bottom of the LPB that lead to a
decrease in the occupancy of states with low wavevectors $k$ with
respect to the occupancy of high-$k$ states. This, in particular,
hinders the polariton system from reaching BEC.

The most straightforward method to study MC polariton relaxation is
measuring the dynamics of the MC emission under optical pumping by
short pulses
\cite{Sermage,Muller,Klopotowski,Renucci,Bajoni1,Bloch,Martin,Muller2,Alexandrou,Valle,Deng,Erland,Ballarini,Bilykh,Bilykh1}.
The dynamics of the MC emission at different angles of observation
$\Theta$ with respect to the sample normal is of particular interest
as it yields information about filling of states corresponding to
wavevectors $k = E(k) \sin \Theta / \hbar c$, where $E(k)$ is the
energy of observed photons and $c$ is the speed of light. The
results of different studies on this issue are somewhat
contradictory. In \cite{Muller} it was shown for a CdTe MC that, at
sufficiently large negative photon--exciton detunings $\Delta$
($\Delta=E_{\rm ph}(k=0)-E_{\rm ex}(k=0)$, where $E_{\rm ph}(k=0)$
and $E_{\rm ex}(k=0)$ are the energies at minima of the photon and
exciton dispersion curves, respectively), the decay time $\tau$ of
the emission from high-$k$ states is reduced considerably with
respect to the decay time of the emission from the bottom of the
LPB. These results were not confirmed in \cite{Klopotowski}, also
carried out with a CdTe MC; here, $\tau$ was observed to be
independent of $\Theta$ within the accuracy of the experiment. In a
study of a GaAs MC \cite{Bajoni1}, it was found that the emission
from $k \approx 0$ decays twice as fast as the emission from the
high-$k$ states. The effect was explained under the assumption that
relaxation to the states with $k \approx 0$ proceeds via
polariton--polariton scattering and, thus, its rate depends
quadratically on the occupancy of the reservoir of exciton-like
states. On the other hand, the high-$k$ polariton states are close
in energy to the reservoir and their occupancy directly reflects the
reservoir depletion.

Qualitative disagreement between the results mentioned above
indicates that the mechanism of filling of radiative MC states is
not universal, but depends on such factors as the excitation
density, photon--exciton detuning etc. In the present study, we
measure the dynamics of the MC emission as a function of the angle
of observation $\Theta$ and the excitation density at a large
negative photon--exciton detuning. It is shown that at high
excitation densities the decay time of the LPB emission increases
with $\Theta$, in agreement with the results of \cite{Bajoni1}. At
low excitation densities $\tau$ is almost independent of $\Theta$.

In order to obtain further information on the mechanism responsible
for the population of states with $k \approx 0$, we also study the
effect of additional continuous-wave (CW) laser illumination on the
MC emission dynamics. The additional illumination was used in a
number of studies for creating free electrons \cite{Bajoni1, Ramon1,
Tartakovskii1, Perrin} (the polariton--electron scattering is
believed to accelerate the relaxation substantially \cite{Malpuech})
and heating up the exciton system \cite{Amo1}. The differential
technique used in the present study enables us to measure changes in
the MC emission dynamics caused by the additional CW illumination.
As a result, it has been shown that the additional illumination
leads to the acceleration of relaxation to the $k \approx 0$ states,
which is associated with an increase in the electron--hole pair
number; this process has a nonlinear character.

\section{Experimental details}
The sample under study was a $\frac{3}{2}\lambda$ MC with mirrors
made of alternating AlAs and Al$_{0.13}$Ga$_{0.87}$As layers. Two
stacks of three tunnel-isolated In$_{0.06}$Ga$_{0.94}$As quantum
wells (QW) were embedded into the GaAs cavity at the positions of
the two electric-field antinodes of the MC. The cavity was grown
wedge shaped, allowing to change the photon mode energy by moving
the excitation spot along the sample surface while keeping the
exciton-mode energy unchanged. The Rabi splitting for the sample was
equal to 6 meV. The sample was mounted inside a variable-temperature
helium cryostat. All the experiments were done at a temperature of
10 K and photon--exciton detuning $\Delta = -5$~meV.

The sample was excited by the emission of a mode-locked Ti-sapphire
laser generating a periodic train of 2.5-ps-long pulses at a
repetition rate of 76~MHz. The excitation energy (1.595~eV;
wavelength 777.6~nm) was larger than the GaAs bandgap (1.519~eV),
which determines the height of the barriers for the QWs. The angle
of incidence of the excitation laser beam was $60^{0}$ with respect
to the sample normal.

Spectrally-resolved dynamics at different angles of observation was
measured using a monochromator coupled to a Hamamatsu C5680 streak
camera. The photoluminescence (PL) from the sample after passing
through a diaphragm with an aperture of $2^0$, was coupled by a lens
into an optical fiber, whose opposite end was fixed at the
monochromator slit. The end of the optical fiber to which the PL is
collected was mounted, together with the diaphragm and the lens, on
a rotation rail allowing registration of the PL coming out at a
given angle. The time and spectral resolution of this system were
30~ps and 0.3~meV, respectively.

In the second part of this study, the sample was illuminated by a CW
532-nm second-harmonic emission of a Nd:YVO$_4$ laser in addition to
the pulsed excitation. The two laser beams were symmetric with
respect to the sample normal and excited the same spot on the sample
at an angle of $60^{0}$. The PL coming out at normal direction with
respect to the sample plane within a $2^{0}$ aperture was
registered.

In these experiments, the PL dynamics was measured using the
picosecond up-conversion optical gating technique \cite{Shah}. The
up-converted (sum-frequency) emission, generated by mixing the PL
with a delayed gate pulse (split from the excitation laser beam) in
a nonlinear crystal, was analyzed by a double monochromator coupled
to a PMT operating in the photon counting mode, and its intensity
was recorded as a function of the delay between the excitation and
the gate pulses. The additional CW illumination was modulated using
a mechanical chopper operating at frequency of 400~Hz and
synchronized with the PMT control unit. With such a setup, both the
dynamics of the main signal (the PL caused by the pulse excitation
only) and the dynamics of the differential signal (the difference
between the PL excited by the pulse excitation with and without
additional CW illumination) were measured. The time and spectral
resolution of this setup were 2.5~ps and 2~meV, respectively.

\section{Microcavity emission dynamics: angular dependence}

\begin{figure}
\begin{center}
\includegraphics[width=\textwidth]{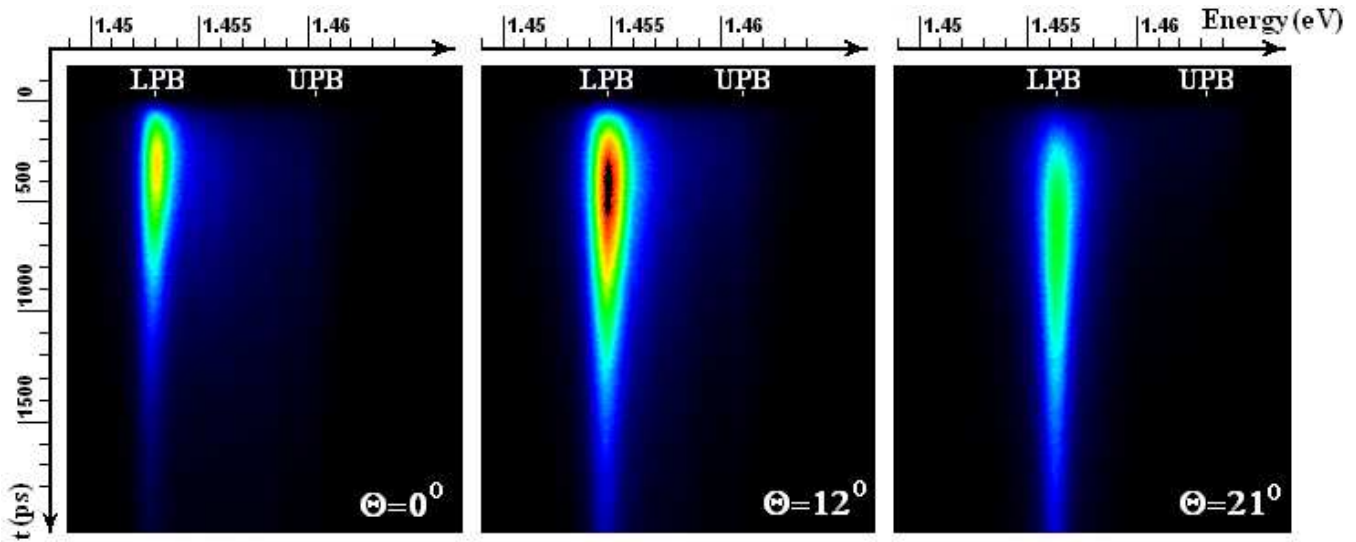}
\caption{Streak camera images of the MC emission collected at
different angles with respect to the sample normal. The positions of
LPB and UPB corresponding to the given angles are indicated. The
time-average excitation density $P = 25$~W/cm$^2$.} \label{Streak}
\end{center}
\end{figure}

The time--spectral dynamics of the MC emission measured at different
angles with respect to the sample normal are shown in
figure~\ref{Streak}. The intense low energy line in the spectra
corresponds to the LPB emission, while the high energy line, which
is much weaker but still recognizable in all spectra, corresponds to
the UPB emission. Lines between the LPB and UPB are attributed to
localized excitons and to scattered LPB emission corresponding to
other angles of observation.

\begin{figure}
\begin{center}
\includegraphics[width=\textwidth]{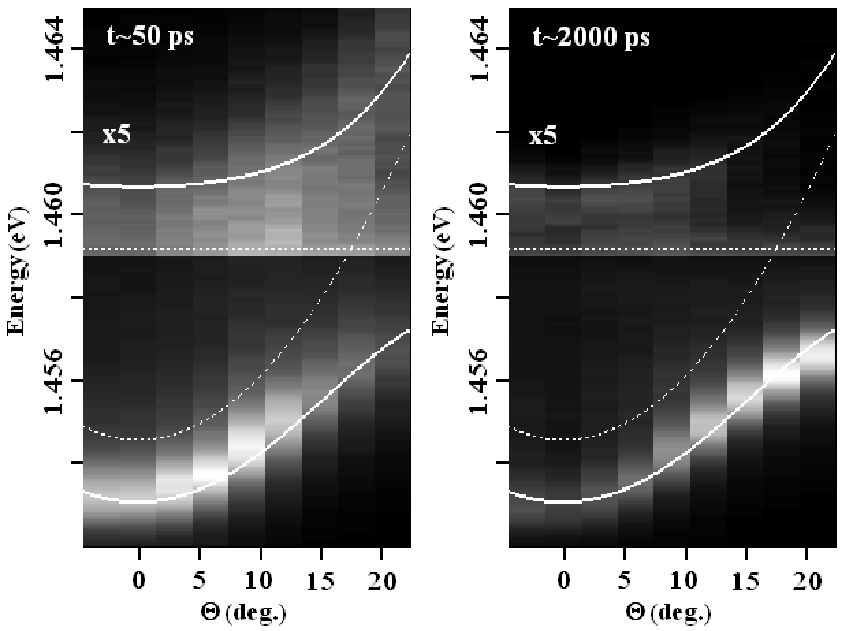}
\caption{Two-dimensional plots of the MC spectra taken at different
angles of observation in the time ranges from 0 to 100~ps (left
image) and from 1950 to 2050~ps (right image) after the excitation
pulse. Lines show the calculated LPB and UPB (solid) together with
the photon and exciton modes (dashed). The intensity for the
emission at photon energies above 1.459~eV is multiplied by a factor
of $5$. The time-average excitation density $P = 25$~W/cm$^2$.}
\label{Disp}
\end{center}
\end{figure}
In order to analyze the dispersion curves for the MC radiative
states, spectra taken at different angles of observation were
combined to yield a series of two-dimensional plots for different
times after the excitation pulse (figure~\ref{Disp}). Solid lines in
figure~\ref{Disp} are polariton branches calculated to fit
experimental data at large time delay from the excitation pulse ($t
\approx 2000$~ps) and dashed lines are corresponding exciton and
photon modes. At the very beginning of the relaxation process ($t
\approx 50$~ps, left image), when the number of electron--hole pairs
in the system is maximal, the LPB lines maxima are blueshifted by
$\sim 0.3$~meV with respect to their position at longer times ($t
\approx 2000$~ps, right image). Nevertheless, the exciton--photon
system is strong-coupled at all times for the given excitation
density $P = 25$~W/cm$^2$.

It should be pointed out that, for $t \approx 2000$~ps
(figure~\ref{Disp}, right image), the bottleneck effect is
pronounced: the low-energy states are considerably depleted with
respect to the high-energy ones. On the other hand, for $t \approx
50$~ps, the emission intensity is larger for the low-energy states.
This fact, however, does not mean a full bottleneck suppression,
because the observed intensity from a given LPB state is related to
its filling factor $f$ by the relation $I = C^{2}f / \tau_{\rm c}$,
where $\tau_{\rm c}$ is the photon lifetime inside the MC and $C$ is
the photon Hopfield coefficient (weight of the photon contribution
in the polariton wave function), which decreases with the angle
$\Theta$ (polariton becomes exciton-like).

\begin{figure}
\begin{center}
\includegraphics[width=\textwidth]{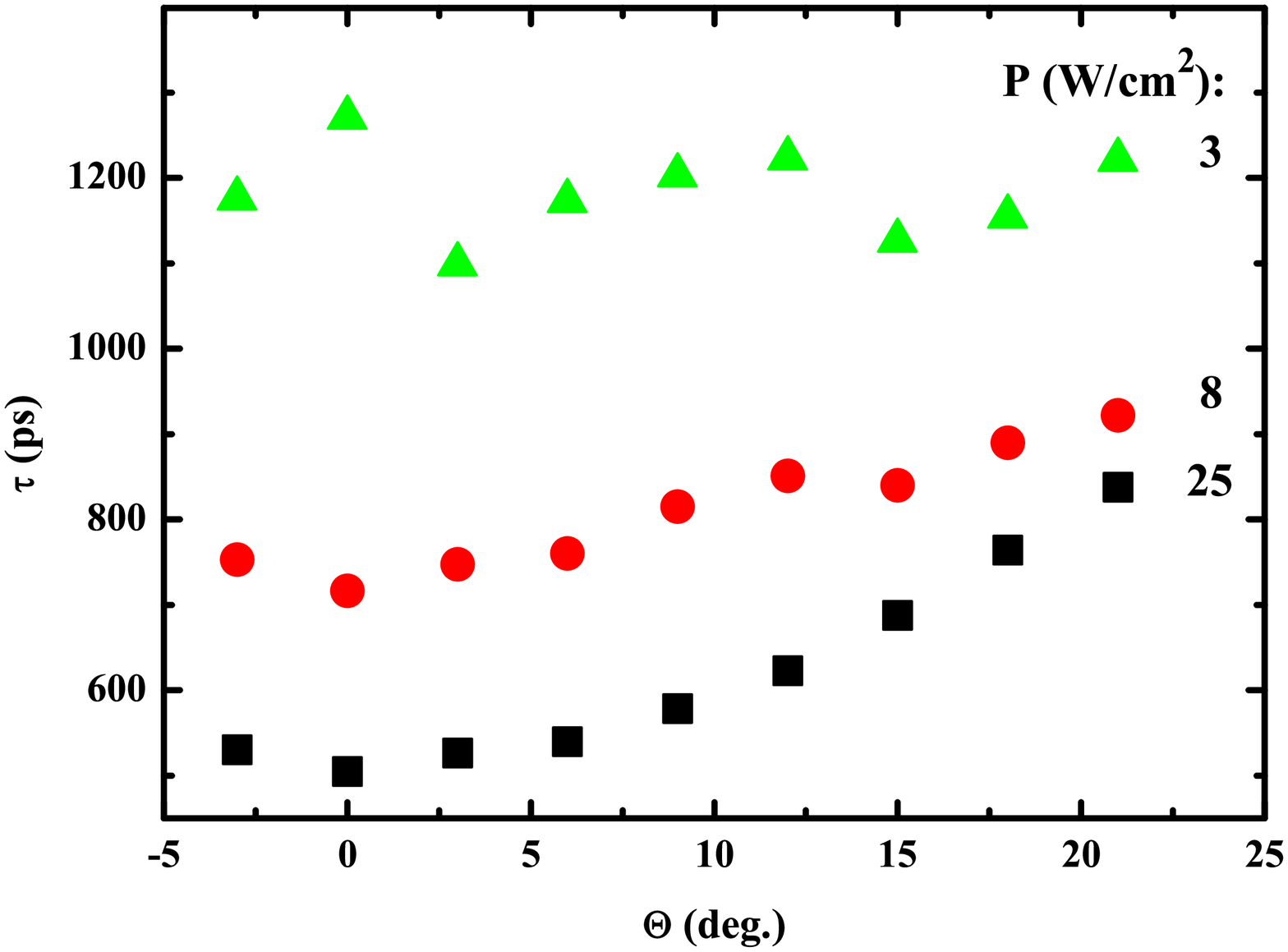}
\caption{The dependence of the LPB emission decay time on the angle
of observation for different time-average excitation densities $P$.}
\label{tauTheta}
\end{center}
\end{figure}
It is seen from figure~\ref{Streak} that the LPB emission dynamics
becomes slower with increasing angle of observation. The angle
dependence of the LPB emission decay time $\tau$ for different
average excitation densities $P$ is presented in
figure~\ref{tauTheta}. One can see that for $P = 25$~W/cm$^2$ $\tau$
increases noticeably with the angle. A similar behavior was observed
in \cite{Bajoni1}, where the dynamics of the emission from the
bottom of LPB was compared to the dynamics of the emission
registered at large $\Theta$. For $P = 8$~W/cm$^2$ the dependence
becomes more flat, and for $P = 3$~W/cm$^2$ $\tau$ no longer depends
on $\Theta$ within the experimental accuracy.

To explain the experimenal results we assume that the LPB emission
dynamics is determined by the depletion rate of the reservoir of
high-$k$ exciton-like states. At sufficiently high excitation
densities the exciton scattering to polariton states with $k \approx
0$ is assisted by collisions with other excitons and remaining
unbound free carriers, whereas the rate of polariton--acoustic
phonon scattering is rather low due to the steep LPB dispersion in
the low-k region (which, for large negative detunings, results in a
large energy difference between the initial and final states)
\cite{Tassone1}. The importance of the interparticle scattering in
polariton relaxation towards the low-k states, leading to the
suppression of the bottleneck effect at high excitation densities,
was previously demonstrated in the CW experiments
\cite{Tartakovskii2, Senellart}. Owing to short polariton lifetimes
at the bottom of the LPB in comparison to typical relaxation times,
the emission intensity $I$ is proportional to the relaxation rate,
which is quadratic in the total number of electron--hole pairs $n$
(in the case of the polariton--polariton and polariton--free carrier
scattering, further referred to as interparticle scattering). Thus,
$I(t) \sim n(t)^{2}$ (strictly speaking, $I(t) \sim w_{\rm ex}(t)
n_{\rm ex}(t)^{2} + w_{\rm e}(t) n_{\rm ex}(t) n_{\rm e}(t) + w_{\rm
h}(t) n_{\rm ex}(t) n_{\rm h}(t)$, where $n_{\rm ex}$, $n_{\rm e}$
and $n_{\rm h}$ are respectively the exciton, electron and hole
densities, $n_{\rm e} = n_{\rm h}$; $w_{\rm ex}$, $w_{\rm e}$ and
$w_{\rm e}$ are respectively the probabilities of the
polariton--polariton, polariton--electron and polariton--hole
scattering \cite{Bajoni1,Bilykh}). Provided the reservoir depletion
follows the law $n(t) = n_{0} \exp(-t/\tau_{\rm x})$, the emission
decays as $I(t) \sim n_{0}^{2} \exp(-2 t/\tau_{\rm x})$, and its
lifetime equals $\tau_{\rm x}/2$. With increasing angle of
observation (polariton wavevector), scattering of polaritons by
acoustic phonons becomes more efficient (because of a decrease in
the energy difference between the final and initial states of
scattered particles), and this process governs the LPB emission
dynamics at sufficiently large angles. For the polariton--phonon
scattering, $I(t) \sim n(t) = n_{0} \exp(-t/\tau_{\rm x})$; thus,
the emission decays with time $\tau_{\rm x}$.

According to figure~\ref{tauTheta}, for $P = 25$~W/cm$^2$ the decay
time $\tau$ decreases from $840$~ps at $\Theta = 21^0$ to $500$~ps
at $\Theta = 0^0$, which indicates a pronounced effect of
interparticle scattering on state filling at the bottom of the LPB.
It should be noted that $\tau$ was determined according to the
emission decay rate in the time interval from 1000 to 2000~ps, when
the number of electron--hole pairs is several times lower as
compared to its initial value. Thus, at the beginning of the
relaxation process, the interparticle scattering might be important
even at low excitation densities.

Another factor contributing to an increase in $\tau$ for states
corresponding to large $\Theta$ might be scattering of polaritons
back to the reservoir, which increases coupling of these states to
the reservoir. Due to this process, $\tau$ becomes closer to
$\tau_{\rm x}$. The rate of backscattering associated with
exciton-like polaritons and free carriers can be estimated from the
linewidth dependence on the excitation density $P$. An increase in
$P$ from $3$ to $25$~W/cm$^2$ leads to an increase in the LPB
linewidth for $\Theta = 21^0$ and $t = 1000$~ps by $\Delta \gamma =
0.2$~meV. This value should be compared with the LPB line broadening
associated with the photon escape from the cavity
$\tilde{\gamma_{\rm c}} = \gamma_{\rm c} C^2$, where $\gamma_{\rm
c}$ is the broadening for a bare photon in the MC. For $\Theta =
21^0$, $\tilde{\gamma_{\rm c}} = 0.5$~meV $> \Delta \gamma$, which
indicates that states corresponding to $\Theta = 21^0$ are not fully
coupled to the reservoir and polaritons mainly escape from these
states via photon emission. Thus, for $P = 25$~W/cm$^2$, scattering
of polaritons back to the reservoir caused by interparticle
collisions cannot fully account for the the observed nearly twofold
increase in $\tau$, which means that the phonon scattering mechanism
discussed above should be involved.

Thus, an increase in the emission decay time with the angle of
observation is presumably related to the crossover of the polariton
states filling mechanism from the interparticle scattering at small
angles to the polariton--acoustic phonon scattering at large angles.

\section{The effect of the additional illumination on the microcavity emission dynamics}
In order to confirm that filling of states with $k \approx 0$ is
controlled by the interparticle scattering mechanism, the effect of
additional nonresonant CW laser illumination on the MC emission
dynamics was studied. The kinetics of the differential signal was
measured, which shows the effect of additional illumination on the
relaxation dynamics of particles created by the laser pulse. If the
additional illumination does not affect this dynamics, the
differential signal simply corresponds to the MC emission created by
the additional illumination alone and does not depend on time. The
additional illumination power density $P_{\rm cw}$ was comparable to
the time-averaged pulse excitation density $P_{\rm pulse}$. Note
that, under these conditions, density of electron--hole pairs
created by the excitation pulse $n_{\rm pulse}= P_{\rm pulse}/(h \nu
F)$ ($F = 76$~MHz is the pulse repetition rate, $h \nu$ is the
excitation photon energy, which in our simple evaluations is assumed
to be approximately the same for the CW illumination and pulse
excitation) is considerably larger than the density of
electron--hole pairs created by the CW illumination $n_{\rm cw} =
P_{\rm cw} \tau_{\rm x} / h \nu$ (where the reservoir depletion time
$\tau_{\rm x} \approx 1$~ns):
\begin{equation}
\frac{n_{\rm cw}}{n_{\rm pulse}} = \tau_{\rm x} F \frac{P_{\rm
cw}}{P_{\rm pulse}}, \label{ncwnpulse}
\end{equation}
which for $P_{\rm cw} = P_{\rm pulse}$ yields  $n_{\rm cw} = F
\tau_{\rm x} n_{\rm pulse} \approx 0.08 n_{\rm pulse}$. Thus, the
additional illumination represents a small perturbation in the first
1--2~ns of the system relaxation. We also note that the photon
energy of the additional CW illumination (2.331~eV) was larger than
the band gaps for the MC-mirror materials (2.24~eV and 1.68~eV for
AlAs and Al$_{0.13}$Ga$_{0.87}$As, respectively) and only the small
portion of the original excitation would reach the cavity.
Newetherless, the time-integrated PL intensity created by the CW
illumination and corresponding to the localized excitons (between
LPB and UPB), which is linear in the excitation power, was close to
that created by the pulse excitation (with photon energy 1.595~eV)
of the same time-averaged power density. Thus, the additional
illumination excites the QWs inside the cavity presumably by the
secondary emission created by recombination of electrons and holes
inside the MC mirrors with almost the same efficiency as the pulse
excitation.

\begin{figure}
\begin{center}
\includegraphics[width=\textwidth]{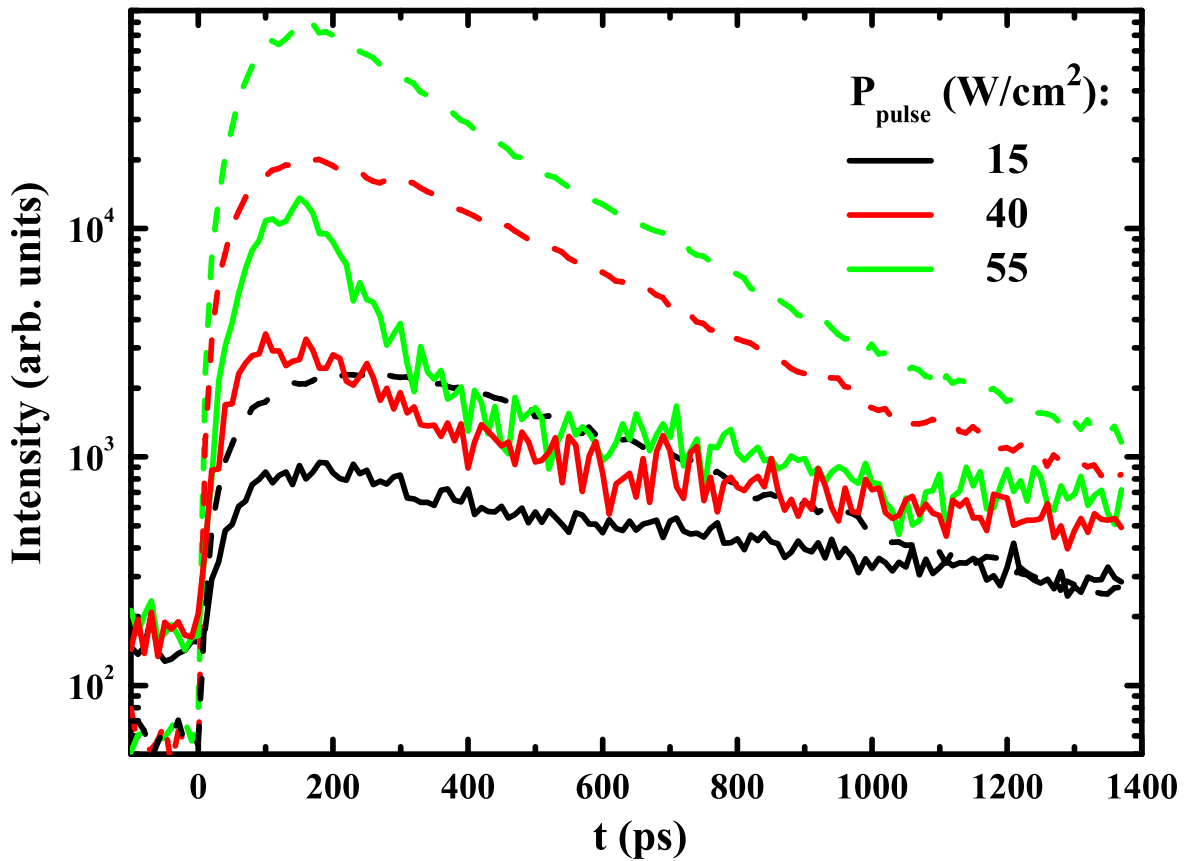}
\caption{Kinetic dependences of the MC emission without additional
CW illumination (dashed lines) and the differential signal (solid
lines) for different time-average pulse excitation densities $P_{\rm
pulse}$ (as indicated in the figure). The CW illumination density
$P_{\rm cw} = 50$~W/cm$^2$.} \label{Illum}
\end{center}
\end{figure}

Figure~\ref{Illum} shows the kinetic dependences of the MC emission
without the additional CW illumination, further referred to as the
main signal (dashed lines), and the differential signal (solid
lines). The following features should be noted. (i) The intensity of
the PL excited by the laser pulses increases under an additional CW
illumination, and this increase (the differential signal at $t > 0$)
is considerably larger than the intensity of the emission excited by
the CW illumination alone (the differential signal at $t < 0$). (ii)
At $t \gtrsim 400$~ps the differential signal decays noticeably
slower than the main signal, the decay rate differing by a factor of
two. (iii) As the pulse excitation density is increased, there
appears a fast component in the differential signal kinetics (at $t
\lesssim 400$~ps), which is presumably related to the onset of
stimulated emission (as the filling factor of the $k \approx 0$
states approaches unity). Note that the observed stimulation is
accompanied by the transition of the system to the weak coupling
regime \cite{Bilykh,Bilykh1} and is not related to the polariton
BEC.

Features (i) and (ii) point at the nonlinear character of the
emission from the bottom of the LPB, confirming that interparticle
scattering is responsible for the filling of $k \approx 0$ states.
Under these conditions, the intensity of emission in the absence of
CW illumination varies as
\begin{equation}
I_{\rm pulse}(t) = \alpha n_{\rm pulse}^{2}(t), \label{Ipulse}
\end{equation}
where, for simplicity, the coefficient $\alpha$ is assumed to be
time-independent. The intensity of the differential signal equals
approximately
\begin{equation}
I_{\rm diff}(t) = \alpha (n_{\rm pulse}(t) + n_{\rm cw}(t))^{2} -
\alpha n_{\rm pulse}(t)^{2} \approx 2 \alpha n_{\rm cw}(t) n_{\rm
pulse}(t) \label{Icw}.
\end{equation}
Assuming that the number of particles created by the CW illumination
does not depend on time, we obtain for the intensity decay $I_{\rm
pulse}(t) \sim n_{0}^{2} \exp(-2 t/\tau_{\rm x})$ and $I_{\rm
diff}(t) \sim 2 n_{\rm cw} n_{0} \exp(-t / \tau_{\rm x})$. The above
relations explain the twofold increase in the decay time of the
differential signal as compared to the emission created by the pulse
excitation alone.

\begin{figure}
\begin{center}
\includegraphics[width=\textwidth]{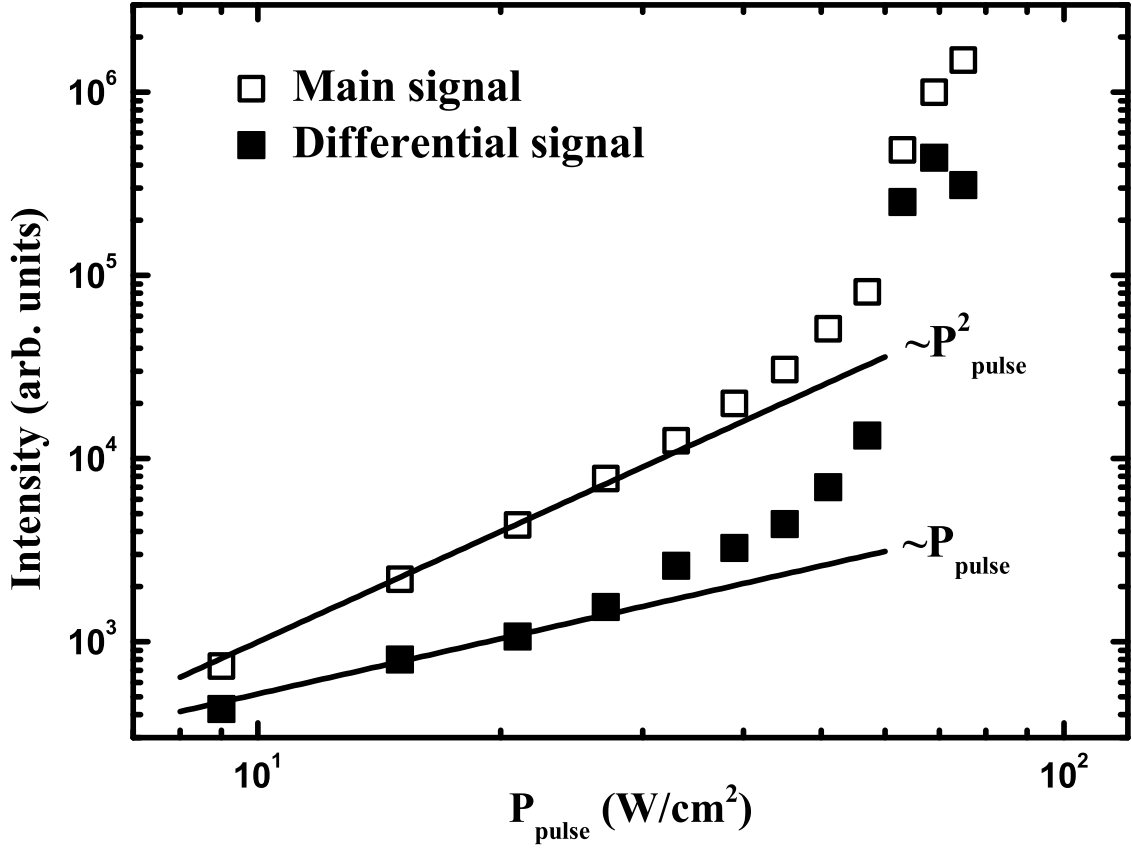}
\caption{Peak intensity dependences for the main signal (open
squares) and differential signal (full squares) on the time-averaged
pulse excitation density $P_{\rm pulse}$. The power density of
additional CW illumination $P_{\rm cw} = 50$~W/cm$^2$.}
\label{Max-P}
\end{center}
\end{figure}

The agreement between the description provided by formulas
(\ref{Ipulse}) and (\ref{Icw}) and the experimental data is also
demonstrated by the dependences of the peak intensities of the main
signal (open squares in figure~\ref{Max-P}) and the differential
signal (full squares in figure~\ref{Max-P}) on the time-averaged
pulse excitation density $P_{\rm pulse}$. It follows from
figure~\ref{Max-P} that at low $P_{\rm pulse}$ the peak intensity of
the main signal is proportional to $P_{\rm pulse}^{2}$, while the
peak intensity of the differential signal is proportional to $P_{\rm
pulse}$ (because $n_0 \sim P_{\rm pulse}$). As $P_{\rm pulse}$ is
further increased, these dependences become steeper. For the peak
intensity of the differential signal, deviation from a linear
dependence, which is related to the fast component in the kinetics
of $I_{\rm diff}$ (figure~\ref{Illum}), becomes apparent at lower
excitation densities $P_{\rm pulse}$ than the fast increase of the
main signal peak intensity begins. That is, the onset of stimulated
emission is more pronounced in the differential signal than in the
main signal. For $P_{\rm pulse} \approx 60$~W/cm$^{2}$, both
intensities increase dramatically (figure~\ref{Max-P}) due to the
onset of lasing in the weak coupling regime \cite{Bilykh,Bilykh1}.

\begin{figure}
\begin{center}
\includegraphics[width=\textwidth]{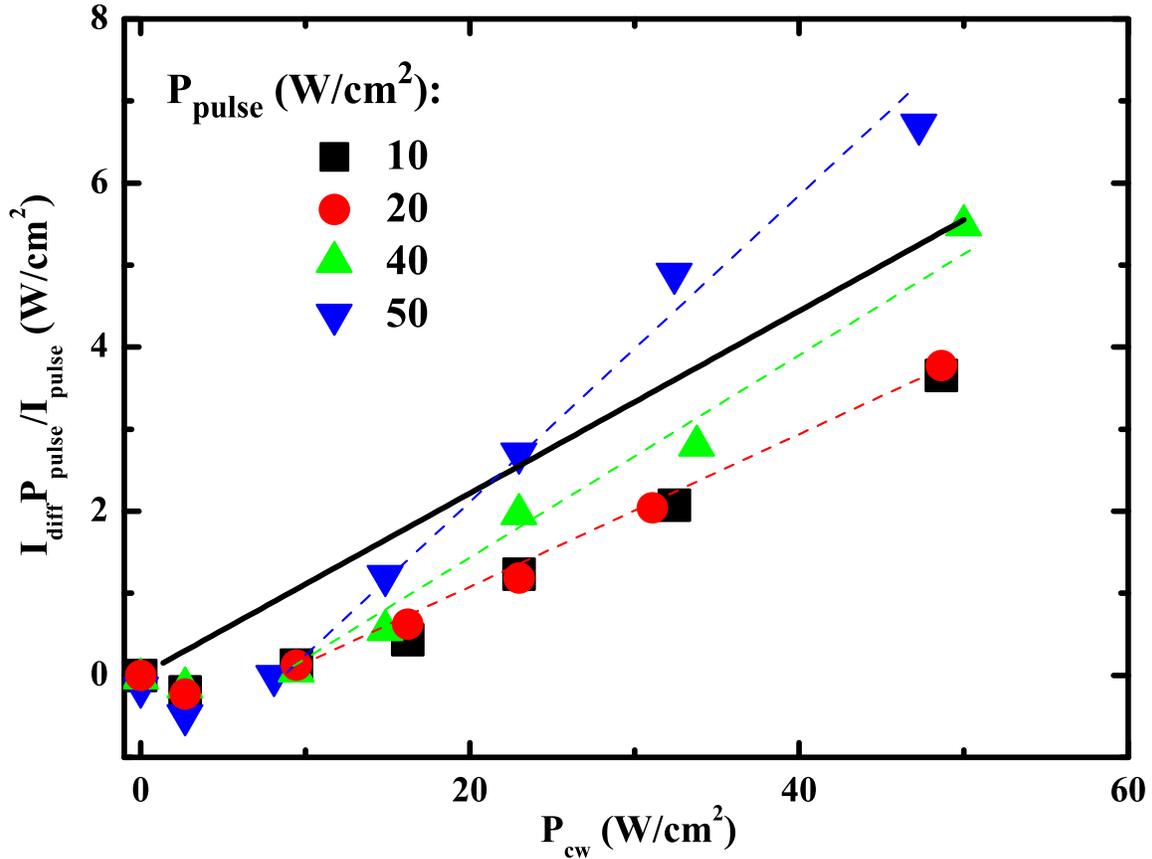}
\caption{Dependences of the normalized differential signal intensity
(see the text) on the power density $P_{\rm cw}$ of the additional
CW illumination for different densities of the pulse excitation
$P_{\rm pulse}$. The calculated dependence is shown by the solid
line; dashed lines are guides to the eye.} \label{Diff-cw}
\end{center}
\end{figure}
Figure~\ref{Diff-cw} shows the peak intensity of the differential
signal (with respect to the differential signal intensity at $t <
0$) as a function of $P_{\rm cw}$ for different values of $P_{\rm
pulse}$. To simplify the experimental data interpretation, the
differential signal intensity was normalized by $I_{\rm
pulse}(t_{\rm peak}) / P_{\rm pulse}$. Thus, the quantity plotted in
figure~\ref{Diff-cw} is $\tilde{I}_{\rm diff} = I_{\rm diff}(t_{\rm
peak}) P_{\rm pulse} / I_{\rm pulse}(t_{\rm peak})$. The data show
that for small CW illumination densities ($P_{\rm cw} \leq
7$~W/cm$^2$), the differential signal intensity is close to zero or
even negative for all pulse excitation levels. At higher $P_{\rm
cw}$, the dependence $\tilde{I}_{\rm diff}(P_{\rm cw})$ becomes
linear and can be described by the following equation:
$\tilde{I}_{\rm diff} = \beta(P_{\rm cw} - P_0)$, where
$P_0=7$~W/cm$^2$. The coefficient $\beta \approx 0.13$ for small
pulse excitation densities $P_{\rm pulse}$ and increases as $P_{\rm
pulse}$ approaches the lasing threshold value. Meanwhile, according
to (\ref{ncwnpulse})--(\ref{Icw})
\begin{equation}
\tilde{I}_{\rm diff} = I_{\rm diff}(t_{\rm peak})\frac{P_{\rm
pulse}}{I_{\rm pulse}(t_{\rm peak})} \approx 2 \tau_{\rm x} F P_{\rm
cw}; \label{Pcw}
\end{equation}
this dependence is shown in figure~\ref{Diff-cw} by a solid line for
$\tau_{\rm x} = 1$~ns. In general, a linear dependence
$\tilde{I}_{\rm diff} = \beta P_{\rm cw}$ is a consequence of the
condition $n_{\rm cw} \ll n_{\rm pulse}$, while the coefficient
$\beta$ is determined by a particular dependence $I_{\rm
pulse}(n_{\rm pulse})$. However, the threshold $P_0$ in the
dependence of the differential signal on $P_{\rm cw}$, the existence
of which is evident in the data, does not appear in (\ref{Pcw}). The
origin of this threshold is not understood at the moment.

Thus, under the given experimental conditions, the main effect of
the additional CW illumination is related to the acceleration of the
relaxation process due to an increase in the number of particles in
the system. At the same time, the possible heating effect of the
additional illumination on the nonequilibrium electron--hole system
\cite{Amo1} seems to be unrelevant. Still, several remarks should be
made. The above conclusions about the significance of the
interparticle scattering in the filling of states at the bottom of
the LPB is valid for the first 1--2~ns only. As the reservoir is
depleted, the efficiency of the interparticle scattering mechanism
decreases and the polariton--acoustic phonon scattering dominates
the filling of $k \approx 0$ states. As a result, the
time-integrated MC emission intensity increases more slowly with the
pulse excitation density than the peak intensity shown in
figure~\ref{Max-P} with open squares.

In the experiments described above one cannot distinguish between
the polariton--polariton and polariton--free carrier scattering
mechanisms. This could be accomplished, for example, by applying an
external magnetic field perpendicular to the MC plane
\cite{Bilykh2}. Preliminary magnetic-field experiments have shown
that, under a CW excitation, the contribution of the
polariton--electron scattering to the total rate of polariton
relaxation to the LPB bottom does not exceed $30\%$ \cite{Belykh}.
As for the results of the present study, it is reasonable to believe
that, at the stage of the emission decay, the majority of charge
carriers are already bound into excitons \cite{Szczytko}, and
filling of $k \approx 0$ states is mainly accompanied by
polariton--polariton scattering. We also note that free electrons
possibly present in the system due to residual doping
\cite{Tartakovskii1} do not contribute significantly to the
polariton relaxation to the $k \approx 0$ states. Otherwise, the
dependence $I_{\rm pulse}(n_{\rm pulse})$ would deviate from a
quadratic one, since scattering by electrons released from
impurities leads to a linear PL intensity dependence on the
excitation power \cite{Bajoni1}, similarly to the case of polariton
relaxation assisted by acoustic phonons.

\section{Conclusions}
The kinetic dependences of the GaAs MC emission from different
states of the LPB have been studied as a function of the pulse
excitation density. Also, the effect of the additional CW
illumination on the MC emission dynamics has been investigated. It
is shown that at low excitation densities the polariton emission
decay rate is almost independent of the polariton wavevector. As the
excitation density is increased, the decay rate of the LPB emission
increases and becomes dependent on the polariton wavevector, so that
it is highest for the emission from the bottom of the LPB and
decreases with increasing polariton energy. The emission intensity
from the bottom of the LPB is nonlinear in the excitation density
and is increased noticeably under the additional CW illumination. On
the basis of the results obtained we conclude that, for a large
negative detuning, filling of states near the bottom of the LPB is
mainly caused by the polariton scattering from the reservoir of
exciton-like states assisted by other exciton-like polaritons and
free charge carriers (except for fairly low excitation densities,
when scattering by phonons becomes important).

\acknowledgements We are grateful to V.D. Kulakovskii, T.V. Murzina
and S.G. Tikhodeev for valuable advice and useful discussions, M.V.
Kochiev and D.A. Mylnikov for help in the experiment. This study was
supported by the Russian Foundation for Basic Research (project
no.~08-02-01438), the Presidium of the Russian Academy of Sciences
(Programs for Fundamental Research) and the Ministry of Education
and Science of the Russian Federation through Federal Program
``Human Resources for Science and Education in Innovative Russia''
(state contract no.~P546). V.V.B. acknowledges financial support
from the Educational and Scientific Center of the P.N. Lebedev
Physical Institute.

\end{document}